\documentclass[journal=jacsat,manuscript=article]{achemso}
\usepackage{graphicx}
\usepackage{bm}
\usepackage{mathptmx}
\usepackage{epstopdf}
\usepackage{booktabs}
\usepackage{multirow}
\usepackage{hhline}
\usepackage{dcolumn}
\usepackage{subfigure}
\usepackage{amsmath}
\usepackage{float}
\usepackage{chemformula}
\usepackage[version=3]{mhchem}
\usepackage{chemfig}
\usepackage{soul}
 
\title{Combining first-principles modeling and symbolic regression for designing efficient single-atom catalysts in Oxygen Evolution Reaction on Mo$_2$CO$_2$ MXenes}

\author{Swetarekha Ram}
\affiliation{Indo-Korea Science and Technology Center (IKST),  Bangalore-560064, India}
\author{Gwan Hyun Choi}
\affiliation{Materials Architecturing Research Center, Korea Institute of Science and Technology,Hwarang-ro 14-gil 5, Seongbuk-gu, Seoul 02792, Republic of Korea}
\author{Albert S. Lee}
\affiliation{Materials Architecturing Research Center, Korea Institute of Science and Technology,Hwarang-ro 14-gil 5, Seongbuk-gu, Seoul 02792, Republic of Korea}
\alsoaffiliation{Convergence Research Center for Solutions to Electromagnetic Interference in future-mobility, Korea Institute of Science and Technology, Hwarang-ro 14-gil5, Seongbuk-gu, Seoul 02792, Republic of Korea}
\author{Seung-Cheol Lee}
\affiliation{Indo-Korea Science and Technology Center (IKST),  Bangalore-560064, India}
\email{leesc@kist.re.kr}
\author{Satadeep Bhattacharjee}
\affiliation{Indo-Korea Science and Technology Center (IKST),  Bangalore-560064, India}
\email{s.bhattacharjee@ikst.res.in}

\begin{document}
\begin{abstract}
 In this study, we address the significant challenge of overcoming limitations in catalytic efficiency for the oxygen evolution reaction (OER). The current linear scaling relationships hinder the optimization of electrocatalytic performance. To tackle this issue, we investigate the potential of designing single-atom catalysts (SACs) on Mo$_2$CO$_2$ MXenes for electrochemical OER using first-principles modeling simulations.
By employing the Electrochemical Step Symmetry Index (ESSI) method, we assess OER intermediates to fine-tune activity and identify the optimal SAC for Mo$_2$CO$_2$ MXenes. Our findings reveal that both Ag and Cu exhibit effectiveness as single atoms for enhancing OER activity on Mo$_2$CO$_2$ MXenes. However, among the 21 chosen transition metals (TMs) in this study, Cu stands out as the best catalyst for tweaking the overpotential ($\eta_{OER}$). This is due to Cu's lowest overpotential compared to other TMs, which makes it more favorable for OER performance. On the other hand, Ag is closely aligned with ESSI=$\eta_{OER}$, making the tuning of its overpotential more challenging. Furthermore, we employ symbolic regression analysis to identify the significant factors that exhibit a correlation with the OER overpotential. By utilizing this approach, we derive mathematical formulas for the overpotential and identify key descriptors that affect catalytic efficiency in electrochemical OER on Mo$_2$CO$_2$ MXenes. This comprehensive investigation not only sheds light on the potential of MXenes in advanced electrocatalytic processes but also highlights the prospect of improved activity and selectivity in OER applications.

\end{abstract}

\section{Introduction}
 Heterogeneous catalysis plays a crucial role in the modern chemical industry, where active sites on the surfaces interact with reactants while the underlying atoms are less involved. Numerous attempts have been made to reduce the size of metal particles in supported metal catalysts used in heterogeneous catalysis. To maximize the utilization of metal atoms in supported metal catalysts, the most effective approach is to reduce metal nanostructures to well-defined metal active sites or single-atom catalysts (SACs) that are \cite{wang2022highly,hu2022atomically} distributed at the atomic level. . The first successful SAC, a single-atom platinum catalyst, showed remarkable results \cite{qiao2011single}. SACs have emerged as a novel class of catalysts with high efficiency and selectivity because of their unique properties, including high surface area, tunable electronic structure, and low coordination number. The feasibility, catalytic activity, empirical achievability, and stability of SACs in chemical processes have been debated for a long time, but advances in modern technologies have made it possible to produce and precisely define SACs~\cite{wang2004dopants,qiao2011single,fu2003active,sun2013single,huang2012catalytically}  catalytically.

Fuel cells and metal-air batteries, for example, are strongly in the focus of green energy technologies due to the increasing energy demand and pollution worldwide.
To develop a clean and sustainable energy infrastructure, renewable power generation technologies such as wind and solar energy are attractive options. For energy conversion and storage technologies such as fuel cells, electrolyzers, and batteries to operate more efficiently, the kinetics of these reactions, including oxygen reduction (ORR), oxygen evolution (OER), hydrogen evolution (HER), and CO2 reduction (CO2RR), must be improved.
Researchers are actively exploring new materials and catalysts to accelerate these reactions and improve the efficiency of these technologies.
\cite{dresselhaus2001alternative,chu2012opportunities}
In terms of availability and sustainability, OER is the most promising candidate.\cite{dau2010mechanism}
For the development of proton exchange membrane (PEM)-based electrolyzers, acidic water electrolysis has a number of advantages over alkaline electrolyzers, including lower ohmic losses, higher voltage efficiency, higher gas purity, more compact system design, higher current density, faster system response, and wider part-load range.\cite{chalenbach2013pressurized}
Iridium (Ir) and ruthenium (Ru) based materials dominate as electrocatalysts for acidic OER.\cite{shi2019robust, reier2017electrocatalytic,pedersen2018operando} The OER is a kinetically slow anodic process that requires a significant overpotential to produce detectable current. Despite the fact that nonprecious metal-based alkaline water electrocatalysts are gaining popularity. Electrocatalytic oxygen evolution is important in the development of hydrogen energy because it encourages the development of novel electrocatalysts and is dedicated to the discovery of materials with high electrocatalytic efficiency.

Several transition metal compounds (TMCs) have been identified as promising bifunctional electrocatalysts, including oxides\cite{lubke2018transition}, sulpides\cite{schmachtenberg2015low}, phosphides\cite{tian2015cobalt}, and hydroxides\cite{li2017directed}.
MXenes, a special type of TMCs with a distinct two-dimensional (2D) structure, have attracted great interest due to their special optical, mechanical, and electrical properties. MXenes have recently attracted attention as potential substrates for a variety of applications, especially in the field of catalysis.
As a promising catalytic system, MXene has been extensively studied in a number of important electrocatalytic processes, including the OER,\cite{kan2020rational}, CO$_2$RR,\cite{li2020synergistic} (HER),\cite{wang2020accelerating} nitrogen reduction reaction (NRR),\cite{huang2019single} oxygen reduction reaction (ORR).\cite{liu2018termination}
However, by integrating SACs into a suitable 2D substrate, which not only increases the catalyst utilisation but also improves the electrochemical properties of the recombinant system, the catalytic properties can be significantly enhanced. \cite{sun2019materials,chen2019computational,li2021direct} SA serves both as an active center and as a cocatalyst to control local surface electronic structures in electrochemical reactions.
Catalytic activity can change when the electronic structure is altered from the original surface.
Because of their highly exposed active cores, single metal atoms are considered an effective technique for maximising the use of catalytic materials.\cite{wang2022highly,hu2022atomically} According to previous studies, the optimal interaction between individual atoms and substrates limits the adsorption of intermediates that can facilitate charge transfer.\cite{ma2021single,wang2022situ} MXenes, in reality, expose the (0001) surface after being etched from the MAX precursor, which is equivalent to the unstable (111) surface in TMCs.\cite{morales2019thickness}
Given that metal cluster formation is limited and single-metal atoms are therefore locked as catalytic active sites, one would wonder if MXenes could be considered of as potential hosts for single-metal atoms.
\cite{wang2018catalysis} 
 We do, however, highlight the absence of thorough studies, and no in-depth comparisons of the 3$d$ and 4$d$ Transition metal (TM) single atom catalysts on Mo$_ 2$CO$_ 2$ MXene for OER have been conducted. 
High durability and activity and strong bonding between metal single atoms and coordination atoms are for improved electrocatalytic performance of SACs.\cite{fei2018microwave,han2018electronic}. 

The aim of this study is to provide the readers with a comprehensive understanding of the crucial role and influence of individual atomic active sites in Mo$_2$CO$_2$ MXene for OER applications and enable significant advances in the field of MXenes. based SACs. We have performed a thorough investigation of the OER activities of 21 individual transition metal atoms on Mo$_2$CO$_2$ using density functional theory (DFT) calculations. We carefully selected and evaluated potential single-atom active sites and also investigated promising anchoring sites. Furthermore, we constructed mathematical formulas for the overpotential through symbolic regression analysis, thereby gaining deeper insight into the key factors affecting the catalytic efficiency in the electrochemical OER on Mo$_2$CO$_2$ MXenes.

\section{Computation Methods}
Using the Vienna Ab Initio Simulation Package (VASP), first-principle spin-polarized simulations are performed. \cite{hafner2008ab}
The PBE functional can be used to describe exchange-correlation effects.
\cite{perdew1996generalized}
The projector augmented wave technique is used to study the electron-ion interactions.
\cite{blochl1994projector}
A plane wave basis set with a kinetic energy cutoff of 500 eV and k-points sampled using a 4 $\times$ 4 $\times$ 1  Monkhorst-Pack mesh are produced in order to optimise the geometry.
A 12 $\times$ 12 $\times$ 1 supercell is used for all electronic calculations. The electronic eigen functions of MXenes are described by the energy convergence $10^{-5}eV $ and the force criterion 0.01 eV\AA$^{-1}$ .
In order to prevent interactions between repeated slabs, at least a 20 {\AA} vacuum layer is placed between each slab along the c direction.
The van der Waals interaction has been taken into consideration using Grimme's semi-empirical DFT-D3 technique. \cite{grimme2006semiempirical} Although the MXene surfaces lack inherent dipoles, the adatom on one side of the slab generates a dipole perpendicular to the surface.
We then employed the dipole correction to remove the artificial field that the periodic boundary conditions imposed on the slab in order to obtain the appropriate adsorption energies.
\cite{bengtsson1999dipole}

The binding energy, $E_b$, for any TM$_{SA}$ that binds to Mo$_2$CO$_2$ may be computed as 
\begin{equation}
      E_{b} = E (TM_{SA}@Mo_2CO_2) - E (Mo_2CO_2) - E(TM_{SA})_{box}
 \end{equation}
 
  where  $E (TM_{SA}@Mo_2CO_2)$, $E (Mo_2CO_2)$, and  $E(TM_{SA})_{box}$ are total energy of the single 3$d$/4$d$ TM adsorbed on Mo$_2$CO$_2$ mono layer, energy of Mo$_2$CO$_2$ mono layer MXene and energy of the single 3$d$/4$d$ TM in vacuum. For all $TM_{SA}@Mo_2CO_2$ scenarios, binding energy is negative, $E_{b}  < 0 $ as shown in Figure-\ref{Fig:str}, and the larger the negative $E_b$ value, the stronger the interaction between Mo$_2$CO$_2$ and the TMs.

 The adsorption energy ($\Delta E_{ads}$) of OER intermediates are calculated according to the following equation
 \begin{equation}
     \Delta E_{ads} = E(ad/TM_{SA}@Mo_2CO_2) - E(TM_{SA}@Mo_2CO_2) - E(ad)
 \end{equation}
 
 where  E(ad), E($TM_{SA}@Mo_2CO_2$) and E(ad/$TM_{SA}@Mo_2CO_2$)  are the total energy of the  free OER intermidiates (O, OH, OOH), the substarte, TM$_{SA}$ attached Mo$_2$CO$_2$ monolayer and 
the TM$_{SA}$ attached Mo$_2$CO$_2$ monolayer  adsorbed with new OER intermidiates (O, OH, OOH), respectively.

   The computational hydrogen electrode concept was used to calculate the free energy diagrams (CHE).\cite{norskov2005trends} In this framework, the chemical potential of gaseous H$_2$ at equilibrium can be used to reference the free energy of the electron-proton pair ($H^+ + e$) (0 V vs standard hydrogen electrode). The following formulas can be used to calculate the change in free energy:

\begin{equation}
\Delta G = \Delta E_{ads} + \Delta E_{ZPE} - T\Delta S  +  \Delta G_U +  \Delta G _{pH}  
\end{equation}

DFT directly calculates the electronic energy difference using $\Delta E_{ads}$. T is the  room temperature (T = 298.15 K), $\Delta E_{ZPE}$ is the change in zero-point energy (ZPE), and $\Delta S$ is the change in entropy. After frequency calculations, the ZPE and vibrational entropy of adsorbed species were determined, and the entropies of gas molecules were taken from reference values.\cite{linstrom2001nist}  The $S$ values of the adsorbed intermediates are
equal to zero and can be neglected according to a previous report.\cite{valdes2008oxidation,rossmeisl2005electrolysis} $\Delta G_U$ is the free energy change with respect to the
applied potential U and the number of varied electrons ($\Delta G_U$ = $-neU$). $ \Delta G _{pH}$ is the correction to the free energy of $H^+$, which can be expressed as
\begin{equation}
  \Delta G_{pH} =  ln10 \times k_BT \times pH  
\end{equation}

 where $k_B$ is the Boltzmann constant. $pH$ was assumed to be zero in this work for the acidic medium.

 The over potential ($\eta_{OER}$) was used to describe the lowest energy requirement for the OER process, which can be calculated as follows
\begin{equation}
\eta_{OER} = {\frac {-\Delta G_{max}}{e} } -  1.23V 
\label{eq:oer}
\end{equation}

\section{Result and discussion}
\subsection{Structure and optimization}
\begin{figure}
  \centering
     \subfigure[]{\includegraphics[scale=0.6]{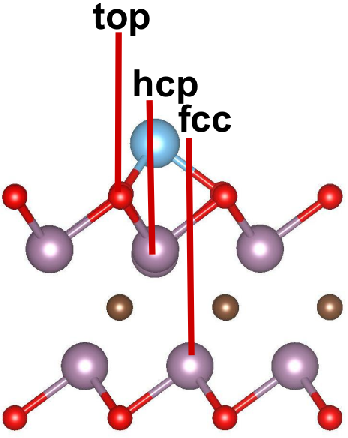}} \hspace{0.2 in}
     \subfigure[]{\includegraphics[scale=0.6]{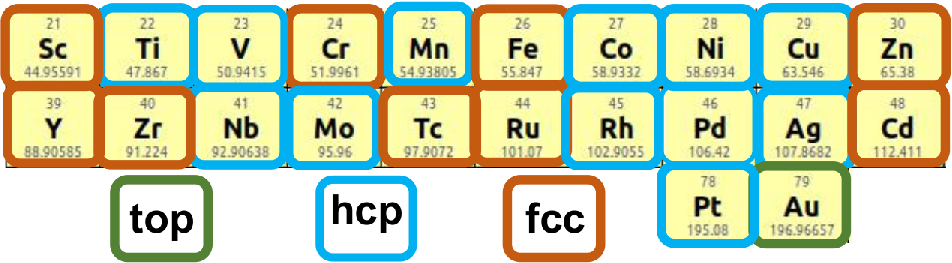}}
 \caption{(a) shows the possible adsorption site for the transition metal, (b) shows all the TM elements included in this work, whereas blue-color square denotes for occupying hcp site, red color for  prefering fcc adsorption site and green colour for top adsorption site. }
 \label{Fig:crystal}
\end{figure}

The structures of monolayer Mo$_2$C with different functional groups were initially our main focus.
According to previous studies,\cite{gao20172d} the surface termination groups attach to the bare MXenes. Due to the exposed metal atoms, which are strong electron donors, there is a strong interaction between the bare MXenes and the functional groups. According to electron energy loss spectroscopy (EELS), the MXene lacks the F group.\cite{sang2016atomic}
In addition, the OH group\cite{xie2014role} would shift to the O group after high-temperature annealing and lithiation, making it possible to experimentally prepare MXenes with O end groups.\cite{zhang2016ti}
After considering all these aspects, we decided to carry out our studies with oxygen-terminated MXenes.
There are several combinations of sites (fcc, hcp, and top) that can be used in an MXene for functional group termination 
. The functional O groups are preferably located at the hcp site on both sides of the Mo$_ 2$C surface, which is consistent with earlier observations.\cite{kuznetsov2019single, gan2020theoretical}
The optimized Mo$_2$CO$_2$ monolayer for a stable configuration has a lattice constant of 2.87 {\AA} which agrees well with previous reports\cite{cheng2018single}. Due to the weak interaction between the O-terminal functional groups and the OER reactive intermediates ($^*$OH, $^*$O and $^*$OOH), it is important to embed active metal atom sites on the surface of Mo$_2$CO$_2$ to enhance the ability to activate the reactive intermediates. From this point of view, the potential adsorption sites for single atoms TM (TM$_{ SA }$) are also considered to determine the energetically favorable adsorption configurations. The selected TM$_{ SA }$ configuration with the lowest energy is chosen for further analysis. The structure of TM$_{ SA }$@Mo$_2$CO$_2$ for the hcp adsorption site of TM is shown in Figure-\ref{Fig:crystal} (a). Figure-\ref{Fig:crystal}(b) shows all TM elements included in this work.

The cohesive energy ($E_{coh}$),  $    E_{coh} = \frac{(E_{TM,bulk} - n E_{TM, box})}{n}$ of the associated bulk TM is taken into consideration to assess the clustering propensity of the adatoms. The supported adatom's tendency to cluster due to the high cohesive energy of the associated bulk metal phase is well recognized to be one of the major barriers to the synthesis of SACs. 
We then define what $E_{diff}$ is, $E_{diff} = E_b - E_{coh}$.
We calculated the difference $E_{diff}$ between the binding energy, ($E_{b}$) of the metal atom on the substrate and the cohesive energy ($E_{coh}$) of the related metal in order to verify that the TM atom on the MXene surface does not aggregate into clusters or nanoparticles.\cite{xia2020atomically} This will ensure that well-dispersed atoms, at least from a thermodynamic perspective, are preferred over metal clusters, which is a requirement that SAC candidates must meet.\cite{cheng2019identification,hu2010density}  

\begin{figure}
  \centering
    \includegraphics[scale=0.6]{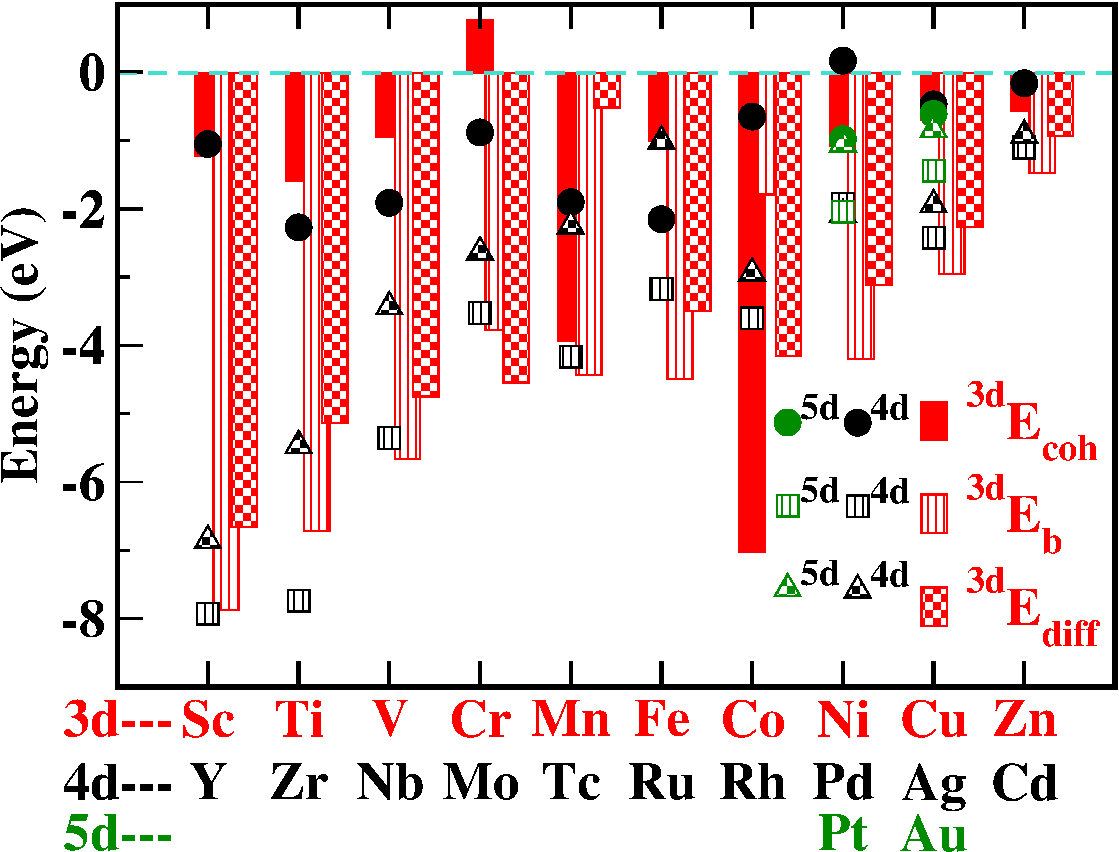} \hspace{0.2 in}
 \caption{The solid filled symbols represent the Cohesive energy, $E_{coh}$, the vertical line filled symbols represent the Binding energy, $E_b$, and the check board filled symbols reflect the difference in Cohesive and Binding energy, $E_{diff}$, which indicates whether the single atom will form a cluster or not. Green is for 5d TM, red for 4d TM and black for 3d TM. }
 \label{Fig:str}
\end{figure}

According to the definition, the values of $E_{diff}$ $<$ 0 or $E_{diff}$ $>$ 0 denote the preference for isolation or clustering of adsorbed single atoms. According to our computational results, shown in Figure-\ref{Fig:str}, all TM$_{ SA }$ studied adsorb to Mo$_2$CO$_2$ instead of clustering. Surprisingly, the 3$d$ and 4$d$ TM follow similar trends for $E_{b}$ and $E_{diff}$ on Mo$_2$CO$_2$ MXenes, but with significant fluctuations that are much larger for the early TMs and become smaller as the series progresses (Figure-\ref{Fig:str}). The exothermic nature of TM on Mo$_ 2$CO$_ 2$ is evident from the negative value of $E_b$, and a structure with a high negative $E_b$ value is more stable. Figure-\ref{Fig:str} and Table S1 show the calculated $E_b$ values. It's interesting to note that for $3d$ and $4d$ metal elements, the binding energies decrease with increasing $d$ electron number.

\subsection{Catalytic activity of single atom  anchored on Mo$_2$CO$_2$}
\subsubsection{Adsorbate evolution mechanism for OER}
\begin{figure}
  \centering
    {\includegraphics[scale=0.6]{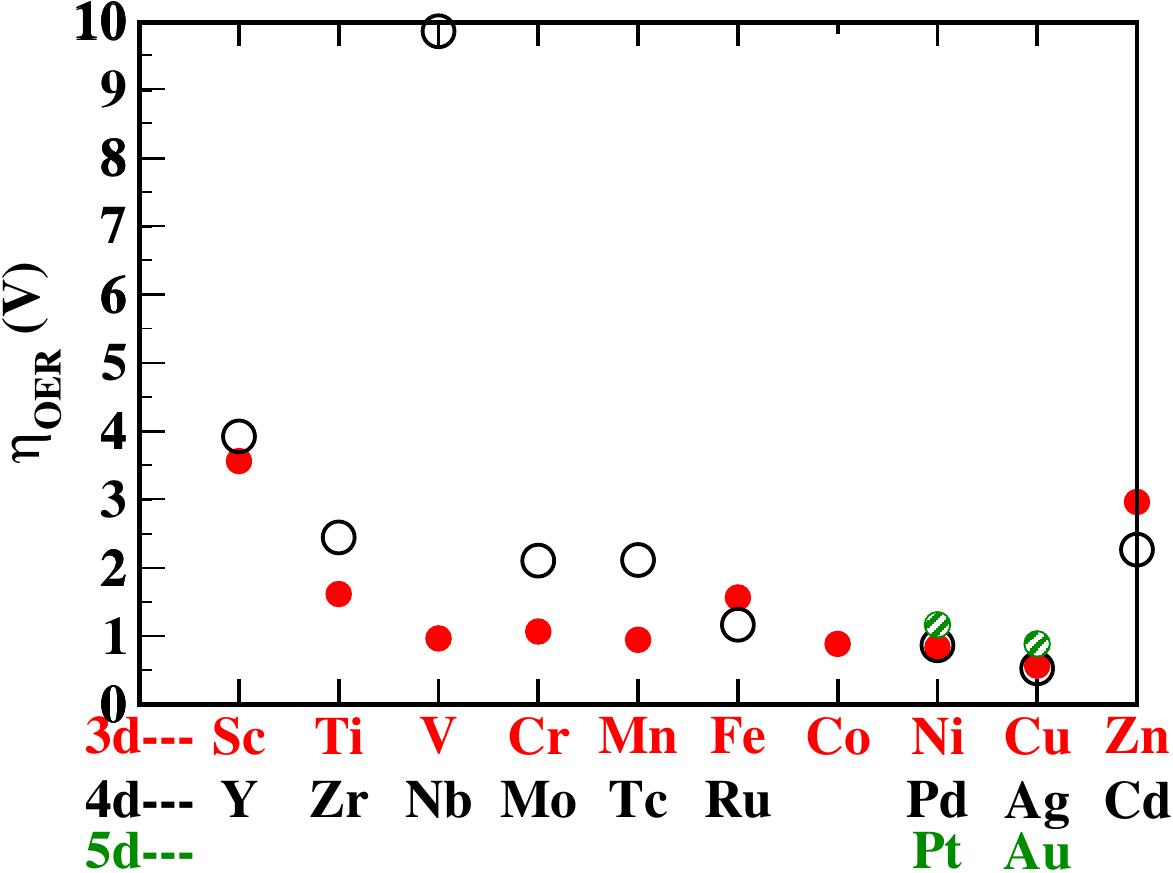}} 
 \caption{The calculated overpotential for selected $3d$, $4d$ and $5d$ TM$_{SA}$ on Mo$_2$CO$_2$ MXene surface. Ag and Cu have least $\eta_{OER}$, 0.53 and 0.57 V respectively among all SAC.}
 \label{Fig:activity}
\end{figure}

The reaction pathways and mechanisms were further explored in accordance with the energetically advantageous adsorption configurations in order to assess the prospective catalytic performance of TM$_{SA}$@MXene monolayers as electrocatalysts. We have studied the catalytic activity of single atoms (3d- TM:-Sc, Ti, V, Cr, Mn, Fe, Co, Ni, Cu and Zn, 4d- TM:- Y, Zr, Nb, Mo, Tc, Ru, Rh, Pd, Ag, Cd and 5d- TM:- Pt, Au) anchored to energetically stable sites of Mo$_2$CO$_2$ by calculating the adsorption energy of OER intermediates (O, OH, OOH) followed by free energy calculation ($\Delta G$). To find the active sites of OER intermediates such as O, OH, and OOH, adsorption on the TM$_{SA}$@Mo$_2$CO$_2$ surfaces was estimated.
Since O terminates the Mo$_2$CO$_2$ surface, the OER species preferentially bind to TM atoms as their primary adsorption sites. The adsorption energy of all OER intermediates at stable sites is shown in Table-S1. Previously, the activity of OER on metal-based catalysts was thought to be a consequence of the difference between the adsorption energies of $O^*$ and $OH^*$(the asterisk denotes the adsorption site). A good catalyst should allow adsorption of species without being too strong or too weak.  
To demonstrate the possibility of using TM$_{SA}$@Mo$_2$CO$_2$ as
OER catalysts, we calculated the overpotential ($\eta_{ OER }$). It is generally accepted that the overpotential ($\eta_{ OER }$), as described in Equation-\ref{eq:oer}, is a useful indicator for evaluating the catalytic activity of the OER catalyst. The $\eta_{ OER }$ of OER on all TM$_{SACs}$ are summarized in Table-S2. The fundamental reaction step with the largest change in free energy is the potential-determining step (PDS), which often requires the least $\eta_{ OER }$ for the reaction and can be the rate-determining step in a catalytic reaction.
\cite{yang2018modulating} A low $\eta_{ OER }$ value indicates better catalytic activity.\cite{zhang2015metal}  We may assume that Ag and Cu@Mo$_2$CO$_2$ have excellent potential as OER catalysts based on the calculated $\eta_{ OER }$ as illustrated in Figure-\ref{Fig:activity}.
The calculated Gibbs free energy profiles show that the PDS for OER over Mo$_2$CO$_2$ and TM (TM = Cr, Mo, Tc, and Ru) is the oxidation of  $O^*$ to $OOH^*$ with a limiting reaction barrier ranging from 2.3 to 3.35 eV , the PDS of Mo$_2$CO$_2$ and Nb is the last step to form the  $O_2$ with a limiting reaction barrier of 11.1 eV and for the remaining TM studied, the PDS is the  oxidation of $OH^*$ to $O^*$ with a limiting reaction barrier ranging from 1.76 to 5.16 eV. 

The quantitative advice on the catalytic performance of SACs in the OER process is provided by the DFT determined overpotential $\eta_{ OER }$ of transition metal SACs, as illustrated in Figure-\ref{Fig:activity}.
We have left out the Nb SA for further debate and analysis since we believe it has a higher overpotential than the other SAs. We only considered 20 TMSA in order to prevent significant errors in the discussion that follows.

\subsubsection{Thermodynamics analysis}
\begin{figure}
  \centering
     \subfigure[]{\includegraphics[scale=0.46]{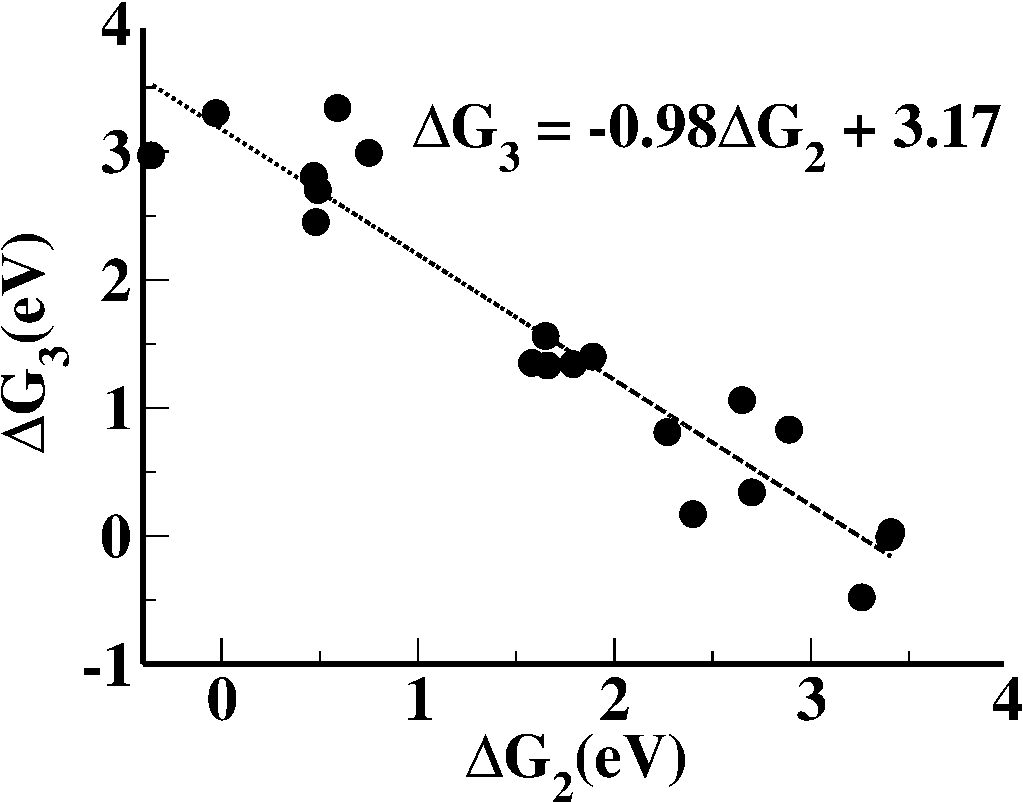}} 
     \subfigure[]{\includegraphics[scale=0.46]{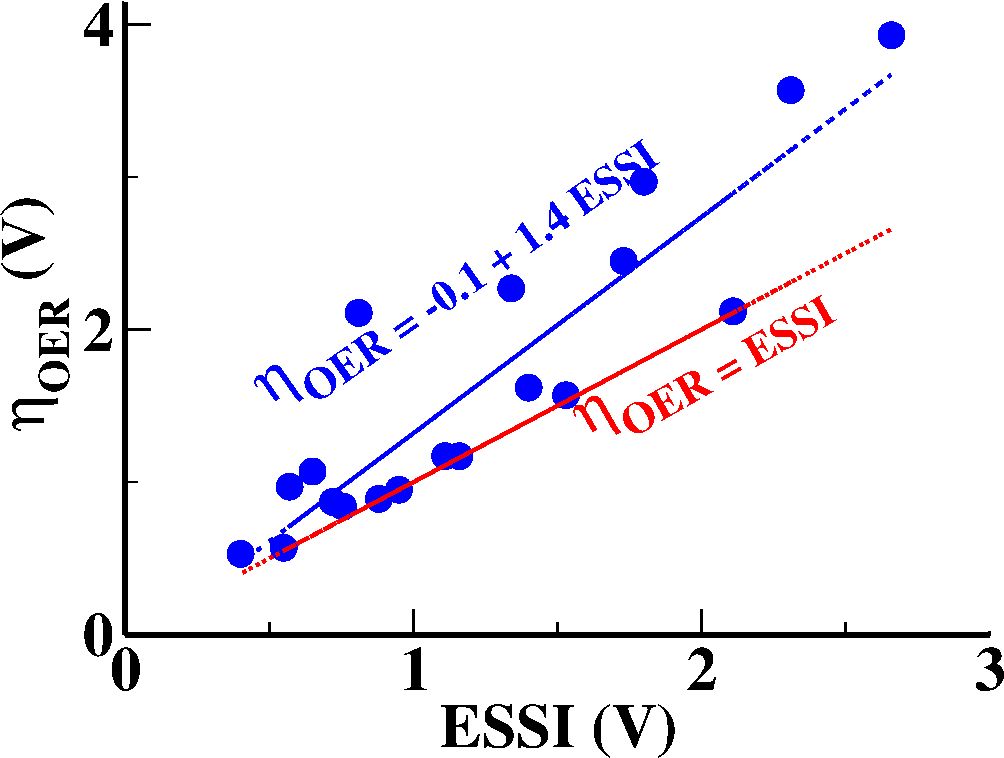}}
     \subfigure[]{\includegraphics[scale=0.6]{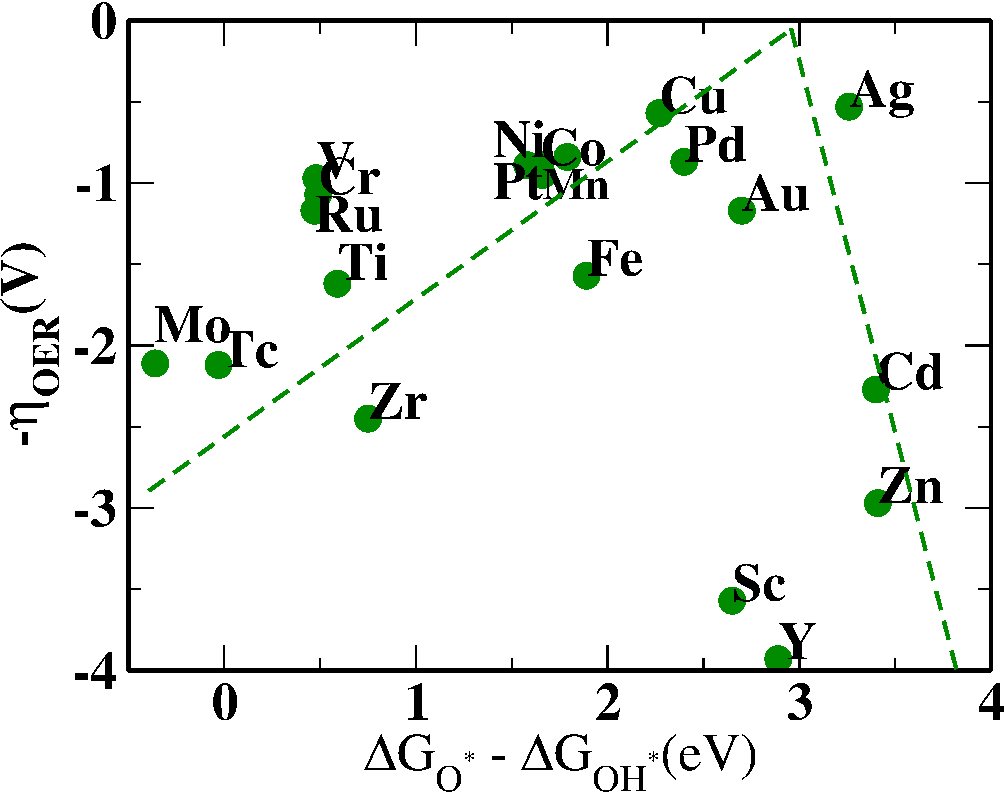}}
 \caption{(a) Linear scaling relation of free energy of OER intermediates (b)$eta_{OER}$ a function of the electrochemical-step symmetry index, ESSI. he red line describes ESSI = $\eta_{OER}$. Materials easy to optimize are on or far from the red line.}
 \label{Fig:ESSI-scaling}
\end{figure}

As showed in Figure-\ref{Fig:ESSI-scaling} (a), we presented the scaling relations between the Gibbs energy of the OER intermediates.
The free-energy changes $\Delta G_3$ ($\Delta G_{OOH^*}- \Delta G_{O^*} $) and $\Delta G_2$ ($\Delta G_{O^*} - \Delta G_{OH^*}$) of the TM$_{SA}$@Mo$_2$CO$_2$ surfaces are scaled based on the computed data, providing a linear scaling relationship (Figure-\ref{Fig:ESSI-scaling} (a)). Linear regression is used to obtain the relevant linear scaling relation in Figure-\ref{Fig:ESSI-scaling} (a):
 $\Delta G_3$ = -0.98  $\Delta G_2$ + 3.17. Since all intermediates have just one top binding accessible, they all have comparable scaling relations.
The obtained linear relation is comparable to prior work, where an offset of 3.2 eV\cite{man2011universality,rossmeisl2005electrolysis} was reported, which is close to our result of 3.17 eV. 
Breaking the scaling relation is an essential but  inadequate prerequisite for optimising OER electrocatalysis.
So we approximated Electrochemical Step Symmetry Index (ESSI)\cite{govindarajan2018does} using equation-\ref{eq:essi} and only applies to steps when $\Delta G_i$ $>$ $E^0$ (denoted $\Delta G_i^+$), as only those can be possible potential determining  steps.
\begin{equation}
    ESSI = \frac{1}{n} \sum_{i=1}^{n} (\Delta G_i^+ - E^0) 
\label{eq:essi}
\end{equation}
More crucially, our projected findings for each SAC satisfy a critical ESSI feature:  ESSI $\le$ $\eta_{OER}$. The same may be seen in Figure-\ref{Fig:ESSI-scaling} (b) that all of the points are either on or above the red line, indicating  ESSI = $\eta_{OER}$. Using the ESSI optimising criteria stated by N. Govindaraja et al.\cite{govindarajan2018does} materials with ESSI = $\eta_{OER}$ have just one step that is bigger than 1.23 eV and are straightforward to optimise since only one step has to be changed to decrease $\eta_{OER}$. Materials close to ESSI  = $\eta_{OER}$ are typically difficult to optimise since they have two or three steps bigger than 1.23 eV and are very comparable in energy, thus changing the potential-determining step may just cause another step to become potential-limiting.  Finally, materials that are somewhat far away from the line where ESSI = $\eta_{OER}$ are excellent candidates for optimisation.  Although they contain numerous steps greater than 1.23 eV, these steps differ significantly, therefore optimising the potential-determining step may result in an overpotential reduction. We screened six $TM_{SA}$,  (Au, Mn, Co, Tc, Pt, Cu and Fe) from the following criteria, ESSI = $\eta_{OER}$ to be the best catalyst for tweaking the overpotential among the 21 chosen TM in this case. Cu is the one with the lowest $\eta_{OER}$, as previously mentioned. At the same time, we realized Ag to be very close to ESSI = $\eta_{OER}$, and as previously noted, tuning the overpotential is quite difficult.

Thus
 the linear scaling relation serves as the beginning point for constructing an overpotential-dependent Volcano plot for the OER over $TM_{SAC}$ to select the best OER activity catalyst. 
 Using the author's recently developed concept of overpotential-dependent volcano plots,\cite{exner2019beyond,man2011universality} we are paying great attention to choosing the best SAC for OER activity. In Figure-\ref{Fig:ESSI-scaling}(c), the volcano plot show the catalytic activity of the investigated TM@Mo$_2$CO$_2$ catalysts.  From the above discussion we conclude Cu and Ag to be a good OER catalyst among all TM SAC studied in the present case, as we find these two TM to very near the peak of the plot with $\eta_{ OER }$ equals to 0.57 and 0.53 V respectively.

 Based on the ESSI scaling relation and volcano plot, we infer that Cu is the best SAC for tuning the over-potential. Nb@Mo$_2$CO$_2$ binds OOH excessively often, causing in the breakdown of O and OH bonds and, eventually, a high $\eta_{ OER }$.

\subsection{Electronic structure}
\begin{figure}
  \centering
    \subfigure[]{\includegraphics[scale=0.6]{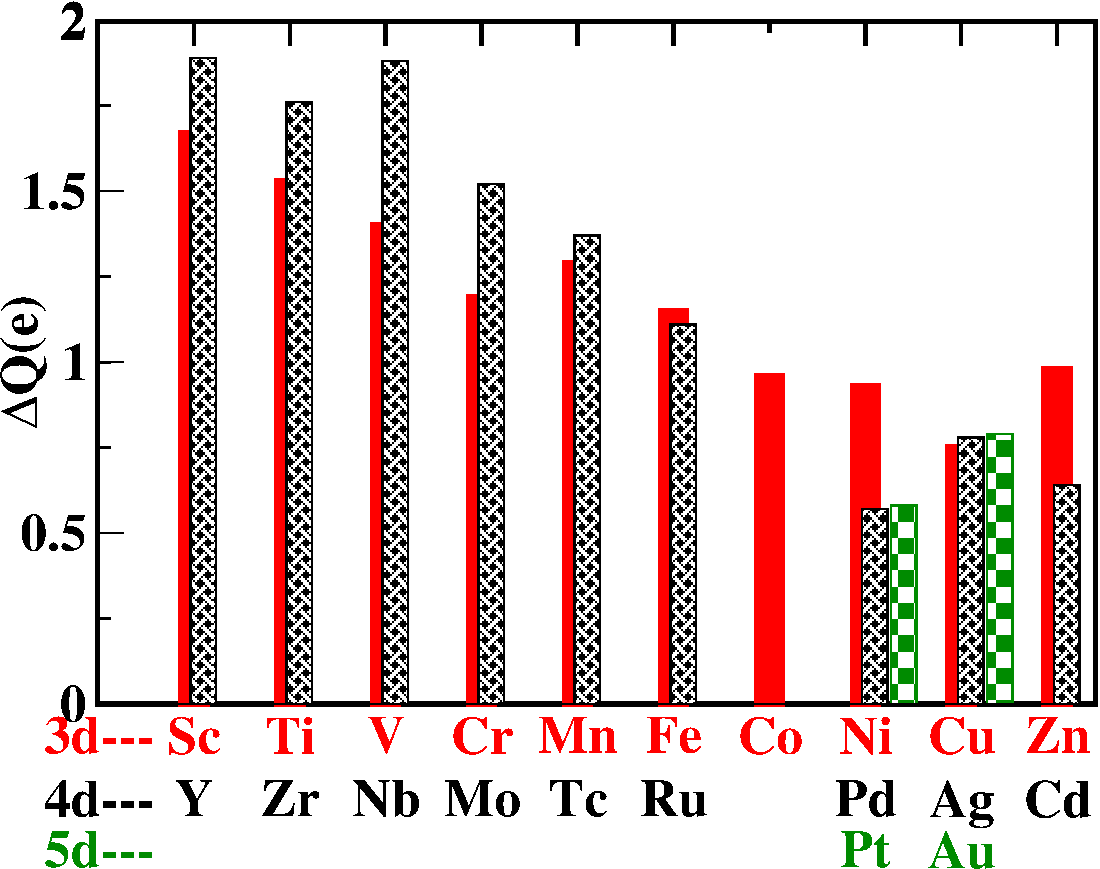} \hspace{0.2 in}}
    \subfigure[]{\includegraphics[scale=0.4]{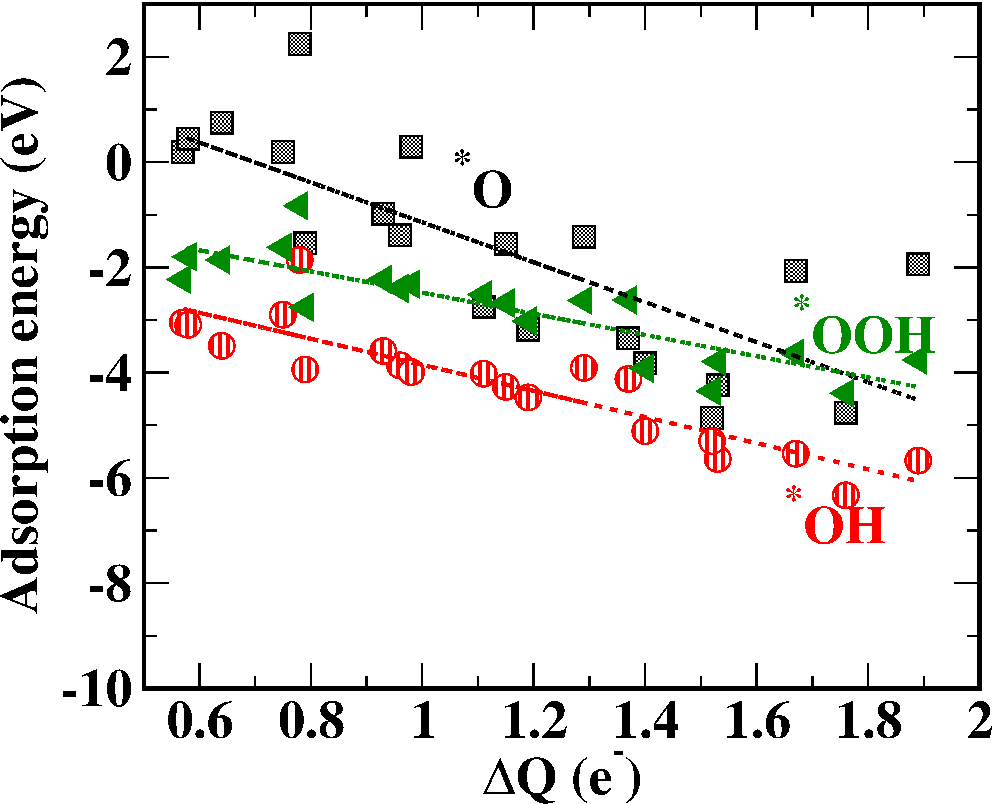} \hspace{0.2 in}}
    \subfigure[]{\includegraphics[scale=0.4]{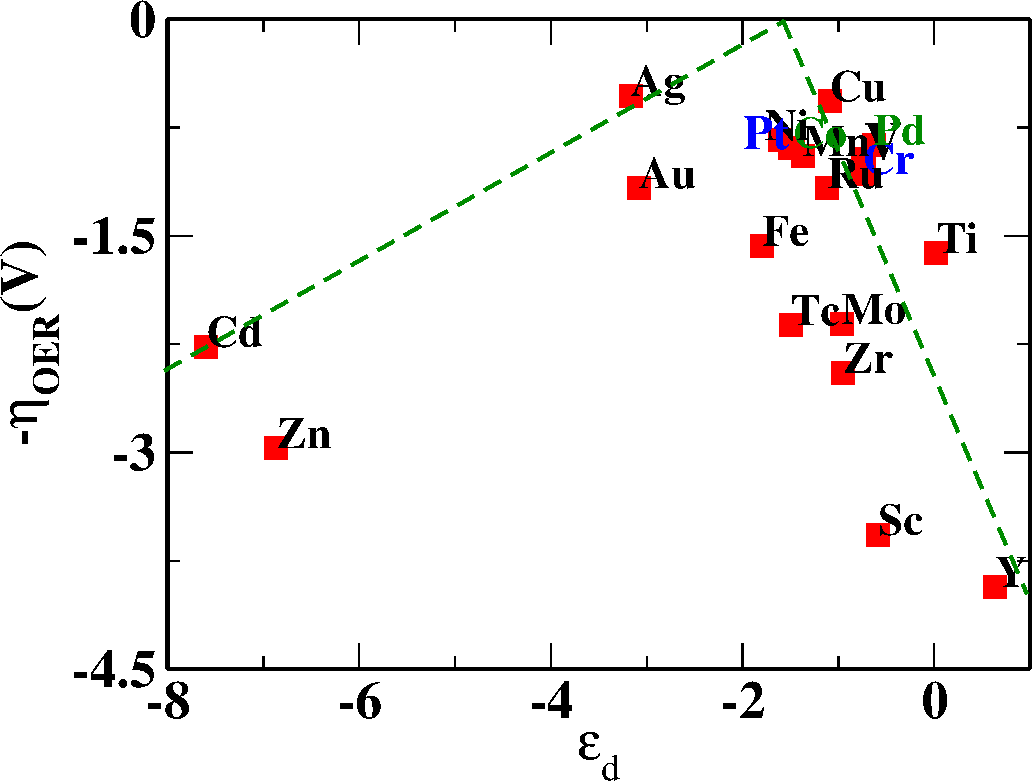}} \hspace{0.2 in}
 \caption{(a)Bar plot for variation of Bader charge on $TM_{SA}$, (b) scaling relation between Bader charge and Adsorption energy intermediates(black for scaling with $^*O$, red for $^*OH$ and green represents for $^*OOH$), (c) Volcano plot for $d$-band center of $TM_{SA}$ and the OER activity, which once again show Ag and Cu to be the best OER catalyst.   }
 \label{Fig:bader}
\end{figure}

 Figure-\ref{Fig:bader} shows the net charge ($\Delta Q$) resulting from electron density analysis using Quantum Theory of Atoms in Molecules (QTAIM) for the adatom developed by Bader.\cite{bader1985atoms} We have shown that ( Figure-\ref{Fig:bader}) all TM components have a tendency to positive charge based on the calculated Bader charge. The formation of SAC is more plausible due to the electrostatic stability caused by the remarkable partial charge transfer from TM to the MXene surface. The results show that the electron density shifts from the TM atom to the nearby O atoms, resulting in a strong bond between Co-O in the TM$_{SA}$@Mo$_2$CO$_2$ monolayer, which is the main requirement for a stable SAC. As the electronegativity of TM increases along the period in the periodic table, the amount of charge transfer to the MXene decreases as one moves down the d series.
 

We examine the electronic structure more closely in order to understand the underlying mechanism by which these SA alter the binding energy. We notice an interesting linear relationship between the Bader charge on the TM and the OER intermediates. Following that, we used the $d$-band centre as a descriptor to examine the OER activity, and we drew a volcano plot for the same $d$-band centre of $TM_{SA}$ with the OER activity as shown in Figure-\ref{Fig:bader} (c). Ag and Cu are once more shown to be the best OER SAC on Mo$_2$CO$_2$ MXene by this. 
\section{Symbolic Regression}
To deepen our comprehension of the essential characteristics that impact overpotential, and to formulate a mathematical expression that integrates these attributes, we executed a symbolic regression analysis~\cite{SR}. We utilized two separate approaches for this purpose, including a genetic algorithm-based method~\cite{GA}and the SISSO (Sure Independence Screening and Sparsifying Operator) technique~\cite{SISSO}. 

Genetic algorithms are often used in traditional symbolic regression. Genetic algorithms are a type of optimization algorithm inspired by the process of natural selection. They generate solutions to optimization problems using techniques inspired by natural evolution, such as inheritance, mutation, selection, and crossover. Symbolic regression uses genetic algorithms to develop mathematical expressions that best fit a given data set. The second approach ~\cite{SISSO} is a different symbolic regression approach that combines it with compressed sensing. Compressed Sensing is a signal processing technique for efficiently acquiring and reconstructing a signal by finding solutions to underconstrained linear systems. This is based on the principle that a signal with a sparse representation can be recovered from a small number of measurements. In the context of data science, or more specifically here, compressed sensing is used in a machine learning environment, specifically for feature selection in symbolic regression. Here the \textit{signal} is the set of potential features (mathematical expressions), and the goal is to select a sparse subset of those features that best corresponds to a given property. Here this is achieved by solving a system of linear equations, a common technique in linear algebra.

Through these methods, we identified the critical factors that affect overpotential, such as Bader charge, electronegativity etc. With the derivation of a mathematical equation that captures the correlation between these variables and overpotential, we can more accurately anticipate and enhance new materials for optimal performance in OER overpotential applications.
\subsubsection{Genetic algorithm}
In this study, we employed a genetic programming-based symbolic regression approach to develop a predictive model from a dataset. We used the \texttt{gplearn} library to create a \texttt{SymbolicRegressor} and integrated custom functions for square and cube operations. The dataset was split into predictor variables ($X$) and the target variable ($y$). We then divided the data into training and testing sets, using an 80-20 split with a random seed of 42.

To create the \texttt{SymbolicRegressor}, we set the following hyperparameters: a population size of 500, a maximum of 20 generations, a stopping criteria of 0.1, and probabilities of 0.5, 0.2, 0.1, and 0.1 for crossover, subtree mutation, hoist mutation, and point mutation, respectively. We also included a parsimony coefficient of 0.0001 to prevent overfitting. The function set was $H^{(m)}\equiv \{(+)(-)(*)(/)(sqrt)(^2)(^3) \}$.

The regressor was fit on the entire dataset ($X$ and $y$), and the model's performance was evaluated using the mean squared error (MSE) between the true target values ($y$) and the predicted values ($\hat{y}$) given by, $\text{RMSE} = \sqrt{\frac{1}{n}\sum_{i=1}^{n}(y_i - \hat{y}_i)^2}
$. 

\subsubsection{Compressed sensing}
The other method that we used for the symbolic regression is Sure Independence Screening and Sparsifying Operator (SISSO), which works in an iterative manner. In this approach, the properties of interest, $P_1$,$P_2$....$P_N$, can be expressed as linear functions of candidate features, ${\bf d}_1$,${\bf d}_2$....${\bf d}_M$.

The model and the descriptors are obtained by minimizing;
\begin{equation}
\underset{C}{\operatorname{arg min}} ||\bf P-\bf D \bf C||^2_2+\lambda||C||_0
\label{CS}
\end{equation}
Where $||\bf C||$ is the $l_0$ norm of c. D is a $N\times M$ matrix known as sensing matrix.
The vector $\bf C$ has $\Omega$-non-zero components known as sparsity. We construct the feature space with the following operators $H^{(m)}\equiv \{(+)(-)(*)(/)(sqrt)(^2)(^3) (exp)(cbrt)\}$ in the case magnetic moment while for the case formation energy we use a large set operators given by $H^{(m)}\equiv\{(+)(-)(*)(/)(sqrt)(^2)(^3)(^-1)\}$.
Within SISSO approach, the candidate features are constructed as non-linear functions of the primary features $\Phi_0$.
The feature space $\Phi_n$ (n=1,2,3..) is made by starting from a set of primary features given by $\Phi_0$ and then combining the features generated in an iterative way using the operators mentioned in $H^{(m)}$.
\begin{figure}
  \centering
    \subfigure[]{\includegraphics[scale=0.65]{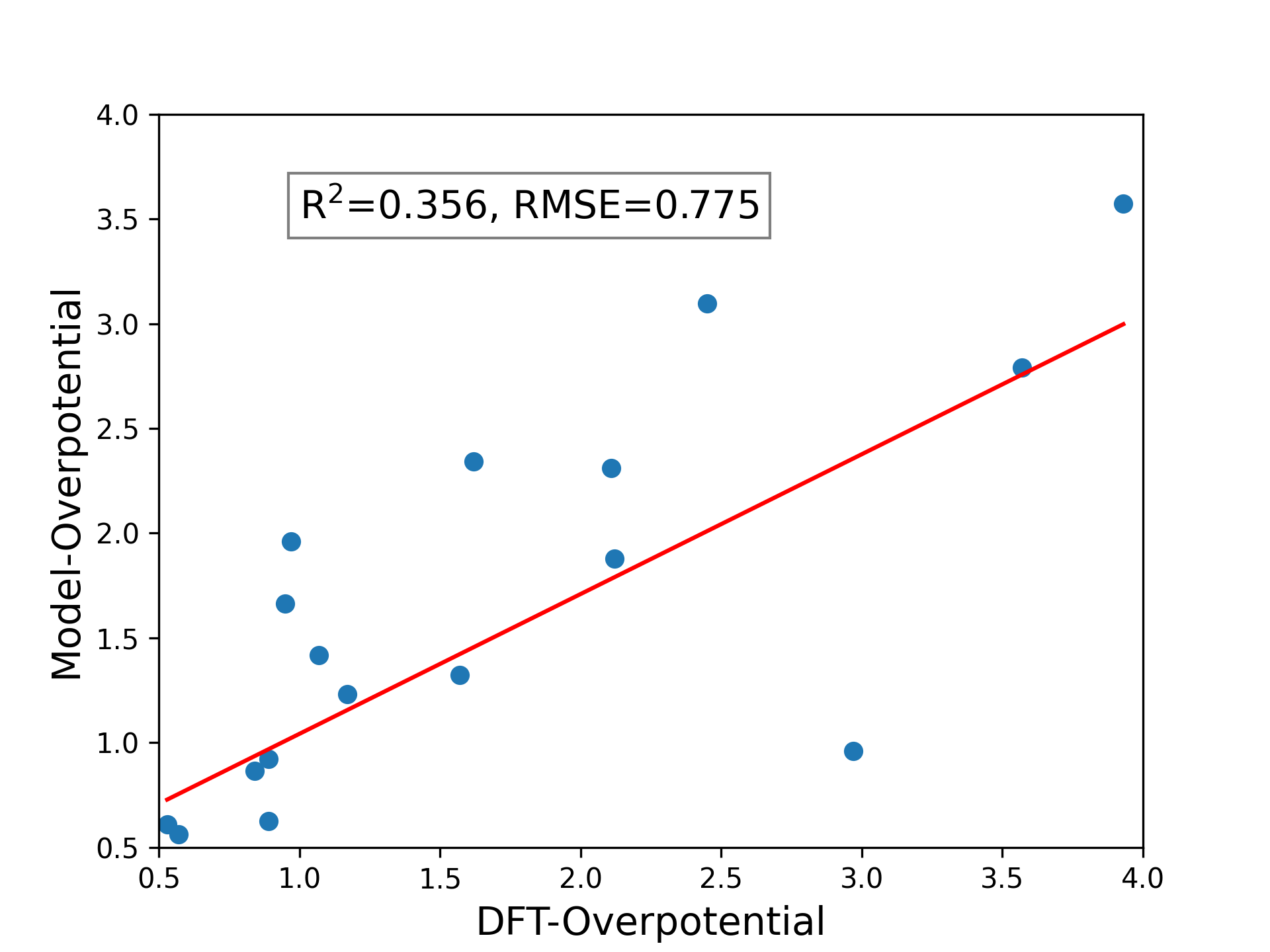} \hspace{0.2 in}}
    \subfigure[]{\includegraphics[scale=0.65]{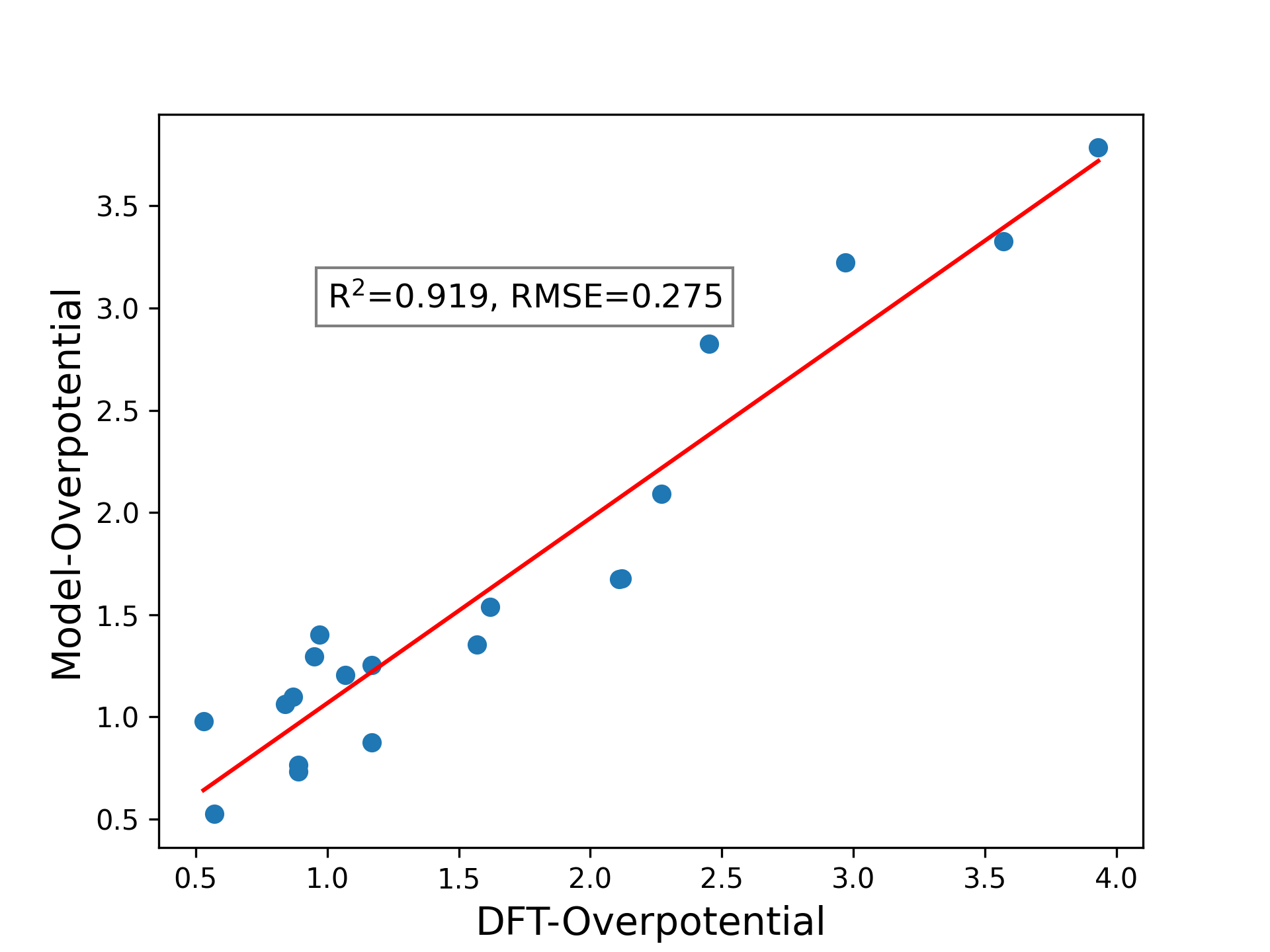}} \hspace{0.2 in}
 \caption{Correlation plot for the symbolic regression study (a) for the genetic algorithm (b) SISSO method.}
 \label{Correlation}
\end{figure}
\subsection{Symbolic regression analysis}
Our study aimed to model the relationship between overpotential and various features, using symbolic regression analysis with two different approaches: genetic algorithm (GA) and compressed sensing (SISSO).

In the Fig.\ref{Correlation}, we show the correlation plot between the model overpotential obtained using GA (top panel) and SISSO (down panel) algorithms with DFT-obtained overpotential. While the GA algorithm gives a somewhat poor correlation ($R^2=0.356$), the root mean square error for GA is 0.775. On the other hand, SISSO provides a very good correlation ($R^2=0.919$) with a root mean square error of 0.275.

Using genetic algorithm, we obtained a simple relationship between overpotential and a single feature, the squared Bader charge ($\eta_{GA}={\Delta Q}^2$). However, when we used compressed sensing with the SISSO code, we discovered a more complex relationship.
The resulting equation for the overpotential using compressed sensing is:
\begin{align}
\eta_{SISSO} &= -0.13\left[ ((\Delta Q\cdot EN)\cdot(E_B+\varepsilon_d)) \right] \\ \nonumber
&\quad -0.043\left[ ((\Delta Q\cdot\varepsilon_d)\cdot(EB-\varepsilon_d)) \right] + 0.081\left[ \left(\frac{WF}{E_{coh}}\right) /(EB+WF) \right] 
\label{ovt}
\end{align}

Here, $EN$ represents electronegativity, $\varepsilon_d$ represents the d-band center, $E_B$ represents binding energy, $WF$ represents work function, $E_{coh}$ represents cohesive energy, and the Bader charge $\Delta Q$ represents the charge transfer between the surface and the single atom catalyst.  
In general, the isolated SACs' unsaturated coordination environment speeds up the electrocatalytic process and increases selectivity\cite{ma2021single}. From our machine learning studies on both GA and the SISSO-based approach, we find that the charge transfer between the MXene and the single atom catalyst is most important. Although the charge transfer is not the sole factor that contributes to the bonding between the TM and the Mo$_2$CO$_2$ MXenes, it does show that this is the primary outcome and is in  perfect agreement with the earlier findings\cite{keyhanian2022effect}.

It should be noted that when comparing the volcano plot obtained in Fig.\ref{Fig:ESSI-scaling} for the OER activity to the volcano plot for standard OER catalysts obtained by Seh \textit{et al.}~\cite{Seh}, the primary difference lies in the role of the substrate. In Seh \textit{et al.}'s work, they also discuss the use of theoretical models, such as volcano plots, to guide catalyst design and optimization based on the differences in Gibbs free energy between O* and OH* adsorption. The paper highlights some of the most promising OER catalysts reported to date, including IrOx/SrIrO3 thin films that exhibit exceptional activity and stability in acid due to their intermediate values of Gibbs free energy difference between O* and OH* adsorption. The key quantity that can change the catalytic behavior, in that case, was the surface area. However, in our case, the main driving phenomenon affecting the OER activity is the coordination number of the single-atom catalyst (SAC) and its interaction with the substrate (MXene) in terms of charge transfer. The substrate also acts as a promoter in this scenario.

Higher $\Delta Q$ means higher overpotential, a trend that is captured by symbolic regression using both the genetic algorithm and SISSO approach (in the equation above, $\varepsilon_d$ is negative, therefore large $\Delta Q$ means large positive $\eta_{SISSO}$). These findings are consistent with the Figure-\ref{Fig:bader} seen previously, where $\Delta Q$ is small for Pt and Ag. A smaller Bader charge implies that there is less energy required to initiate the oxygen evolution reaction (OER), resulting in a smaller overpotential. Therefore, smaller Bader charges are related to smaller overpotentials because they indicate less interaction between the surface and single-atom catalyst, leading to lower energy requirements for OER initiation.
\section{Conclusions}
We thoroughly investigated the OER activities of 21 TM SAs supported on Mo$_2$CO$_2$ using DFT calculations. Potential single atom active sites were selected and evaluated, and prospective anchoring sites were also investigated. Charge transfer from TM to Mo$_2$CO$_2$ causes TM to become an effective OER active site. Moreover, potential possibilities for OER catalysts with high catalytic activity and stability include Cu and Ag@Mo$_2$CO$_2$. The symbolic regression analysis allowed us to establish mathematical formulas for the overpotential, providing a deeper understanding of the key descriptors influencing catalytic efficiency in the electrochemical OER on Mo$_2$CO$_2$ MXenes. In conclusion, our study offers useful theoretical recommendations for building affordable, highly functional OER catalysts.

\section{Acknowledgment}
This work was supported by NRF grant funded by MSIP, Korea 
(No. 2009-0082471 and 
No.
2014R1A2A2A04003865), the Convergence Agenda Program (CAP) of the Korea Research Council of 
Fundamental Science and Technology (KRCF)and GKP (Global Knowledge Platform) project of the 
Ministry of Science, ICT and Future Planning. A.S. Lee acknowledges support by the National Research Council of Science \& Technology (NST) grant by the Korea government (MSIT) (CRC22031-000).
\clearpage
\newpage
\bibliography{TM-SAC}

\end{document}


\begin{table}[]
\begin{tabular}{|llll|llll|llll|}
\hline
\multicolumn{4}{|l|}{3d}                                                                            & \multicolumn{4}{l|}{4d}                                                                            & \multicolumn{4}{l|}{5d}                                                                            \\ \hline
\multicolumn{1}{|l|}{TM} & \multicolumn{1}{l|}{$E_{coh}$} & \multicolumn{1}{l|}{$E_b$} & $E_{diff}$ & \multicolumn{1}{l|}{TM} & \multicolumn{1}{l|}{$E_{coh}$} & \multicolumn{1}{l|}{$E_b$} & $E_{diff}$ & \multicolumn{1}{l|}{TM} & \multicolumn{1}{l|}{$E_{coh}$} & \multicolumn{1}{l|}{$E_b$} & $E_{diff}$ \\ \hline
\multicolumn{1}{|l|}{Sc} & \multicolumn{1}{l|}{-1.21}     & \multicolumn{1}{l|}{-7.87} & -6.66      & \multicolumn{1}{l|}{Y}  & \multicolumn{1}{l|}{-1.06}     & \multicolumn{1}{l|}{-7.93} & -6.86      & \multicolumn{1}{l|}{}   & \multicolumn{1}{l|}{}          & \multicolumn{1}{l|}{}      &            \\ \hline
\multicolumn{1}{|l|}{Ti} & \multicolumn{1}{l|}{-1.58}     & \multicolumn{1}{l|}{-6.71} & -5.13      & \multicolumn{1}{l|}{Zr} & \multicolumn{1}{l|}{-2.28}     & \multicolumn{1}{l|}{-7.74} & -5.47      & \multicolumn{1}{l|}{}   & \multicolumn{1}{l|}{}          & \multicolumn{1}{l|}{}      &            \\ \hline
\multicolumn{1}{|l|}{V}  & \multicolumn{1}{l|}{-0.93}     & \multicolumn{1}{l|}{-5.66} & -4.74      & \multicolumn{1}{l|}{Nb} & \multicolumn{1}{l|}{-1.92}     & \multicolumn{1}{l|}{-5.36} & -3.44      & \multicolumn{1}{l|}{}   & \multicolumn{1}{l|}{}          & \multicolumn{1}{l|}{}      &            \\ \hline
\multicolumn{1}{|l|}{Cr} & \multicolumn{1}{l|}{0.77}      & \multicolumn{1}{l|}{-3.77} & -4.54      & \multicolumn{1}{l|}{Mo} & \multicolumn{1}{l|}{-0.89}     & \multicolumn{1}{l|}{-3.53} & -2.64      & \multicolumn{1}{l|}{}   & \multicolumn{1}{l|}{}          & \multicolumn{1}{l|}{}      &            \\ \hline
\multicolumn{1}{|l|}{Mn} & \multicolumn{1}{l|}{-3.92}     & \multicolumn{1}{l|}{-4.43} & -0.51      & \multicolumn{1}{l|}{Tc} & \multicolumn{1}{l|}{-1.91}     & \multicolumn{1}{l|}{-4.17} & -2.26      & \multicolumn{1}{l|}{}   & \multicolumn{1}{l|}{}          & \multicolumn{1}{l|}{}      &            \\ \hline
\multicolumn{1}{|l|}{Fe} & \multicolumn{1}{l|}{-0.99}     & \multicolumn{1}{l|}{-4.49} & -3.49      & \multicolumn{1}{l|}{Ru} & \multicolumn{1}{l|}{-2.16}     & \multicolumn{1}{l|}{-3.18} & -1.02      & \multicolumn{1}{l|}{}   & \multicolumn{1}{l|}{}          & \multicolumn{1}{l|}{}      &            \\ \hline
\multicolumn{1}{|l|}{Co} & \multicolumn{1}{l|}{-7.02}     & \multicolumn{1}{l|}{-1.78} & -4.15      & \multicolumn{1}{l|}{Rh} & \multicolumn{1}{l|}{-0.66}     & \multicolumn{1}{l|}{-3.61} & -2.95      & \multicolumn{1}{l|}{}   & \multicolumn{1}{l|}{}          & \multicolumn{1}{l|}{}      &            \\ \hline
\multicolumn{1}{|l|}{Ni} & \multicolumn{1}{l|}{-1.07}     & \multicolumn{1}{l|}{-4.19} & -3.11      & \multicolumn{1}{l|}{Pd} & \multicolumn{1}{l|}{0.16}      & \multicolumn{1}{l|}{-1.93} & -2.09      & \multicolumn{1}{l|}{Pt} & \multicolumn{1}{l|}{-0.99}     & \multicolumn{1}{l|}{-2.05} & -1.06      \\ \hline
\multicolumn{1}{|l|}{Cu} & \multicolumn{1}{l|}{-0.69}     & \multicolumn{1}{l|}{-2.95} & -2.26      & \multicolumn{1}{l|}{Ag} & \multicolumn{1}{l|}{-0.48}     & \multicolumn{1}{l|}{-2.43} & -1.95      & \multicolumn{1}{l|}{Au} & \multicolumn{1}{l|}{-0.62}     & \multicolumn{1}{l|}{-1.45} & -0.85      \\ \hline
\multicolumn{1}{|l|}{Zn} & \multicolumn{1}{l|}{-0.55}     & \multicolumn{1}{l|}{-1.47} & -0.92      & \multicolumn{1}{l|}{Cd} & \multicolumn{1}{l|}{-0.17}     & \multicolumn{1}{l|}{-1.11} & -0.93      & \multicolumn{1}{l|}{}   & \multicolumn{1}{l|}{}          & \multicolumn{1}{l|}{}      &            \\ \hline
\end{tabular}
\caption{Cohesive energy,$E_{coh}$, Binding energy, $E_{b}$, Difference of Cohessive and binding energy, $E_{diff}$ for the studied 3$d$, 4$d$ and 5$d$ transition metals are reported here.}
\end{table}

\begin{table}[]
\begin{tabular}{|l|llllll|}
\hline
   & \multicolumn{6}{l|}{3d}                                                                                                                                       \\ \hline \hline
TM & \multicolumn{1}{l|}{$\eta_{OER}$} & \multicolumn{1}{l|}{$\Delta$ G$_1$} & \multicolumn{1}{l|}{$\Delta$ G$_2$} & \multicolumn{1}{l|}{$\Delta$ G$_3$} & \multicolumn{1}{l|}{$\Delta$ G$_4$} & ESSI  \\ \hline
Sc & \multicolumn{1}{l|}{3.57}         & \multicolumn{1}{l|}{-2.14} & \multicolumn{1}{l|}{4.79}  & \multicolumn{1}{l|}{-1.59} & \multicolumn{1}{l|}{2.29}  & 2.31  \\ \hline
Ti & \multicolumn{1}{l|}{1.62}         & \multicolumn{1}{l|}{-2.24} & \multicolumn{1}{l|}{2.83}  & \multicolumn{1}{l|}{2.44}  & \multicolumn{1}{l|}{0.49}  & 1.4   \\ \hline
V  & \multicolumn{1}{l|}{0.97}         & \multicolumn{1}{l|}{-1.71} & \multicolumn{1}{l|}{2.19}  & \multicolumn{1}{l|}{1.97}  & \multicolumn{1}{l|}{1.25}  & 0.57  \\ \hline
Cr & \multicolumn{1}{l|}{1.07}         & \multicolumn{1}{l|}{-1.07} & \multicolumn{1}{l|}{1.56}  & \multicolumn{1}{l|}{2.21}  & \multicolumn{1}{l|}{0.1}   & 0.65  \\ \hline
Mn & \multicolumn{1}{l|}{0.95}         & \multicolumn{1}{l|}{-0.51} & \multicolumn{1}{l|}{2.18}  & \multicolumn{1}{l|}{-0.33} & \multicolumn{1}{l|}{1.1}   & 0.95  \\ \hline
Fe & \multicolumn{1}{l|}{1.57}         & \multicolumn{1}{l|}{-0.87} & \multicolumn{1}{l|}{2.76}  & \multicolumn{1}{l|}{-0.49} & \multicolumn{1}{l|}{1.1}   & 1.53  \\ \hline
Co & \multicolumn{1}{l|}{0.89}         & \multicolumn{1}{l|}{-0.46} & \multicolumn{1}{l|}{2.11}  & \multicolumn{1}{l|}{-0.09} & \multicolumn{1}{l|}{0.62}  & 0.88  \\ \hline
Ni & \multicolumn{1}{l|}{0.84}         & \multicolumn{1}{l|}{-0.19} & \multicolumn{1}{l|}{1.99}  & \multicolumn{1}{l|}{-0.45} & \multicolumn{1}{l|}{0.65}  & 0.756 \\ \hline
Cu & \multicolumn{1}{l|}{0.57}         & \multicolumn{1}{l|}{0.48}  & \multicolumn{1}{l|}{1.78}  & \multicolumn{1}{l|}{-1.46} & \multicolumn{1}{l|}{0.55}  & 0.55  \\ \hline
Zn & \multicolumn{1}{l|}{2.97}         & \multicolumn{1}{l|}{-0.59} & \multicolumn{1}{l|}{4.01}  & \multicolumn{1}{l|}{-3.39} & \multicolumn{1}{l|}{2.04}  & 1.8   \\ \hline  \hline
   & \multicolumn{5}{l|}{4d}                                                                                                                               &       \\ \hline \hline
TM & \multicolumn{1}{l|}{$\eta_{OER}$} & \multicolumn{1}{l|}{$\Delta$ G$_1$} & \multicolumn{1}{l|}{$\Delta$ G$_2$} & \multicolumn{1}{l|}{$\Delta$ G$_3$} & \multicolumn{1}{l|}{$\Delta$ G$_4$} & ESSI  \\ \hline
Y  & \multicolumn{1}{l|}{3.93}         & \multicolumn{1}{l|}{-2.27} & \multicolumn{1}{l|}{5.15}  & \multicolumn{1}{l|}{-2.05} & \multicolumn{1}{l|}{2.64}  & 2.66  \\ \hline
Zr & \multicolumn{1}{l|}{2.45}         & \multicolumn{1}{l|}{-2.92} & \multicolumn{1}{l|}{3.67}  & \multicolumn{1}{l|}{2.25}  & \multicolumn{1}{l|}{1.09}  & 1.73  \\ \hline
Nb & \multicolumn{1}{l|}{9.87}         & \multicolumn{1}{l|}{-2.04} & \multicolumn{1}{l|}{1.84}  & \multicolumn{1}{l|}{-1.76} & \multicolumn{1}{l|}{11.08} & 5.23  \\ \hline
Mo & \multicolumn{1}{l|}{2.11}         & \multicolumn{1}{l|}{-1.9}  & \multicolumn{1}{l|}{1.54}  & \multicolumn{1}{l|}{3.33}  & \multicolumn{1}{l|}{1.25}  & 0.81  \\ \hline
Tc & \multicolumn{1}{l|}{2.12}         & \multicolumn{1}{l|}{-0.72} & \multicolumn{1}{l|}{0.69}  & \multicolumn{1}{l|}{3.34}  & \multicolumn{1}{l|}{-0.93} & 2.11  \\ \hline
Ru & \multicolumn{1}{l|}{1.17}         & \multicolumn{1}{l|}{-0.63} & \multicolumn{1}{l|}{1.1}   & \multicolumn{1}{l|}{2.34}  & \multicolumn{1}{l|}{-0.55} & 1.11  \\ \hline
Pd & \multicolumn{1}{l|}{0.87}         & \multicolumn{1}{l|}{0.34}  & \multicolumn{1}{l|}{2.06}  & \multicolumn{1}{l|}{-2.23} & \multicolumn{1}{l|}{1.83}  & 0.72  \\ \hline
Ag & \multicolumn{1}{l|}{0.53}         & \multicolumn{1}{l|}{1.52}  & \multicolumn{1}{l|}{1.75}  & \multicolumn{1}{l|}{-3.74} & \multicolumn{1}{l|}{1.09}  & 0.4   \\ \hline
Cd & \multicolumn{1}{l|}{2.27}         & \multicolumn{1}{l|}{-0.1}  & \multicolumn{1}{l|}{3.5}   & \multicolumn{1}{l|}{-3.41} & \multicolumn{1}{l|}{1.65}  & 1.34  \\ \hline \hline
   & \multicolumn{5}{l|}{5d}                                                                                                                               &       \\ \hline \hline
TM & \multicolumn{1}{l|}{$\eta_{OER}$} & \multicolumn{1}{l|}{$\Delta$ G$_1$} & \multicolumn{1}{l|}{$\Delta$ G$_2$} & \multicolumn{1}{l|}{$\Delta$ G$_3$} & \multicolumn{1}{l|}{$\Delta$ G$_4$} & ESSI  \\ \hline
Pt & \multicolumn{1}{l|}{0.89}         & \multicolumn{1}{l|}{-0.54} & \multicolumn{1}{l|}{2.11}  & \multicolumn{1}{l|}{-0.22} & \multicolumn{1}{l|}{1.17}  & 0.88  \\ \hline
Au & \multicolumn{1}{l|}{1.17}         & \multicolumn{1}{l|}{0.31}  & \multicolumn{1}{l|}{2.39}  & \multicolumn{1}{l|}{-2.36} & \multicolumn{1}{l|}{1.22}  & 1.16  \\ \hline
\end{tabular}
\caption{Over potential ,$\eta_{OER}$, Free energy difference, $\Delta$G$_1$, $\Delta$G$_2$, $\Delta$G$_3$, $\Delta$G$_4$ and the Electro-chemical step symmetry index, ESSI  for the studied 3$d$, 4$d$ and 5$d$ transition metals are reported here.}
\end{table}